\documentclass[copyright,creativecommons]{eptcs}
\providecommand{\event}{PLACES 2009} 
\providecommand{\volume}{??}
\providecommand{\anno}{2009}
\providecommand{\eid}{??}
\usepackage{amsmath,amssymb,amsthm}
\usepackage{listings}
\usepackage{xspace}
\usepackage{xcolor}
\newtheorem{thm}{Theorem}
\newtheorem{lem}[thm]{Lemma}
\newtheorem{defi}[thm]{Definition}

\newcommand{\etal}{\textit{et al.}\@\xspace}

\newcommand{\ldsq}{[\![} 
\newcommand{\rdsq}{]\!]} 

\newcommand{\procblocktype}[1][\pol x]{(r_2\!: \transtype{\pol x})}

\newcommand{\transtype}[1]{\mathcal{T}\ldsq{#1}\rdsq}

\newcommand{\tagA}[1][H]{\mathcal{A}({#1})}

\newcommand{\tagI}[4]{\mathcal{I}({#1}, {#2}, {#3}, {#4})}
\newcommand{\tagId}[1]{\tagI{#1}\Psi\Gamma\Lambda}
\newcommand{\tagV}[3]{\mathcal{V}({#1}, {#2}, {#3})}
\newcommand{\tagVd}{\tagV v \Psi \Gamma}
\newcommand{\tagT}[1]{\mathcal{T}({#1})}

\newcommand{\tagLock}[3]{{#1}\colon({#2}, {#3})}

\newcommand{\smallrulespc}{-0.1em}
\newcommand{\rulespc}{0.5em}

\newcommand{\erase}{\mathcal{E}}

\newcommand{\milkw}[1]{\textsf{#1}\xspace}
\newcommand{\done}{\milkw{done}}
\newcommand{\malloc}{\milkw{malloc}}
\newcommand{\fork}{\milkw{fork}}
\newcommand{\jump}{\milkw{jump}}
\newcommand{\ifk}{\milkw{if}}

\newcommand{\newlock}{\milkw{newLock}}
\newcommand{\unlock}{\milkw{unlock}}
\newcommand{\tsl}{\milkw{testSetLock}}
\newcommand{\halt}{\milkw{halt}}

\newcommand{\requires}{\milkw{requires}}

\newcommand{\register}[1]{\milkw r_{#1}}

\newcommand{\heapcode}[2]{{#1}\{{#2}\}}
\newcommand{\heapcodev}{\heapcode \tau I}

\newcommand{\heaptuple}[2]{\langle {#1} \rangle^{#2}}
\newcommand{\lockbit}[1]{\textbf{#1}}
\newcommand{\uninit}[1][\tau]{?{#1}} 
\newcommand{\instantiation}[2][v]{{#1}[{#2}]}

\newcommand{\assign}{:=}
\newcommand{\mallocarg}[2]{[{#1}]^{#2}}
\newcommand{\malloci}[1][\mallocarg{\vec\tau}{\lambda}]{r \assign \malloc\ {#1}}
\newcommand{\forki}{\fork\ v}
\newcommand{\cjump}[3]{\ifk\ {#1}={#2}\ \jump\ {#3}}
\newcommand{\cjumpi}[1][v]{\cjump{r}{#1}{v}}
\newcommand{\cjumpli}{\cjumpi[\lockbit 0]}
\newcommand{\movei}{r \assign v}
\newcommand{\loadi}[1][v]{r \assign {#1}[n]}
\newcommand{\storei}[1][v]{r[n] \assign {#1}}
\newcommand{\aopi}[1][r]{r \assign {#1} + v}
\newcommand{\donei}{\done}
\newcommand{\jumpi}[1][v]{\jump\ {#1}}
\newcommand{\newlocki}{\lambda\colon(\Lambda, \Lambda),r \assign \newlock}
\newcommand{\tsli}[1][v]{r \assign \tsl\ {#1}}
\newcommand{\unlocki}{\unlock\ v}

\newcommand{\program}[3]{\langle{#1};{#2};{#3}\rangle}
\newcommand{\processor}[3]{\program{#1}{#2}{#3}}
\newcommand{\poolitem}[2]{\langle{#1},{#2}\rangle}

\newcommand{\numreg}{\milkw R}
\newcommand{\numthr}{\milkw N}

\newcommand{\integer}{\milkw{int}}
\newcommand{\tupletype}[2]{\langle{#1}\rangle^{#2}}
\newcommand{\tupletypet}{\tupletype{\pol\tau}{\lambda}}

\newcommand{\locktype}[1][\lambda]{{#1}}

\newcommand{\subtype}{<:}
\newcommand{\codetype}[2]{{#1}\ \requires\ {#2}}
\newcommand{\codetypeforall}[3]{\forall[{#1}].(\codetype{#2}{#3})}
\newcommand{\codetypet}{\codetype \Gamma \Lambda}
\newcommand{\codetypea}{\codetypeforall {\pol \lambda\colon(\pol
    \Lambda, \pol \Lambda)} \Gamma \Lambda}

\newcommand{\universal}[4]{\forall[{#1}\colon\!({#2},{#3})].{#4}}
\newcommand{\forallt}{\universal\lambda\Lambda\Lambda\tau}

\newcommand{\ftv}{\operatorname{ftv}}
\newcommand{\dom}{\operatorname{dom}}

\newcommand{\Mapsto}{\colon}
\newcommand{\eval}{\operatorname{\hat R}}

\newcommand{\subs}[2]{[{#1}/{#2}]}
\newcommand{\update}[2]{\{{#1}\Mapsto {#2}\}}

\newcommand{\mkRrule}[1]{\textsc{R-#1}\xspace}
\newcommand{\mkTrule}[1]{\textsc{T-#1}\xspace}
\newcommand{\mkSrule}[1]{\textsc{S-#1}\xspace}
\newcommand{\Rhalt}{\mkRrule{halt}}
\newcommand{\Rschedule}{\mkRrule{schedule}}
\newcommand{\Rfork}{\mkRrule{fork}}
\newcommand{\Rjump}{\mkRrule{jump}}
\newcommand{\Rmove}{\mkRrule{move}}
\newcommand{\Rarith}{\mkRrule{arith}}
\newcommand{\RbranchT}{\mkRrule{branchT}}
\newcommand{\RbranchF}{\mkRrule{branchF}}

\newcommand{\Rnewlock}{\mkRrule{newLock}}
\newcommand{\RtslZ}{\mkRrule{tsl}0}
\newcommand{\RtslO}{\mkRrule{tsl}1}
\newcommand{\Runlock}{\mkRrule{unlock}}

\newcommand{\Rmalloc}{\mkRrule{malloc}}
\newcommand{\Rload}{\mkRrule{load}}
\newcommand{\Rstore}{\mkRrule{store}}
\newcommand{\Ttype}{\mkTrule{type}}
\newcommand{\Tint}{\mkTrule{int}}

\newcommand{\Tlabel}{\mkTrule{label}}
\newcommand{\Tlock}{\mkTrule{lock}}
\newcommand{\Tuninit}{\mkTrule{uninit}}

\newcommand{\Treg}{\mkTrule{reg}}
\newcommand{\Tvalapp}{\mkTrule{valApp}}
\newcommand{\Tdone}{\mkTrule{done}}
\newcommand{\Tfork}{\mkTrule{fork}}
\newcommand{\Tnewlock}{\mkTrule{newLock}}
\newcommand{\Ttsl}{\mkTrule{tsl}}
\newcommand{\Tunlock}{\mkTrule{unlock}}
\newcommand{\TbranchC}{\mkTrule{critical}}
\newcommand{\Tmove}{\mkTrule{move}}
\newcommand{\Tarith}{\mkTrule{arith}}
\newcommand{\Tbranch}{\mkTrule{branch}}

\newcommand{\Tmalloc}{\mkTrule{malloc}}
\newcommand{\Tload}{\mkTrule{load}}
\newcommand{\Tstore}{\mkTrule{store}}
\newcommand{\Tjump}{\mkTrule{jump}}

\newcommand{\SregFile}{\mkSrule{regFile}}

\newcommand{\pad}{\;\;}
\newcommand{\Space}[1]{\pad{#1}\pad}
\newcommand{\grmeq}{\Space{::=}}

\newcommand{\grmor}{\;\mid\;}

\newenvironment{myfigure}{
  \begin{figure}[t]\centering}{
  \end{figure}}

\newcommand{\myfbox}[1]{\protect\fbox{#1}}

\newcommand{\myparagraph}[1]{\paragraph{#1}}

\lstdefinestyle{numbers}
    {numbers=left, stepnumber=1, numberstyle=tiny, numberstep=10pt}
\lstdefinestyle{nonumbers}
    {numbers=none}

\lstdefinelanguage{mil}{
  morekeywords={version, registers, new, exists, existsL, unpack,
    pack, packL, as, jump, if,share,
    dones, type, done, lock, testSetLock, unlockE, unlockS, unlock,
    newLock, fork, tslE, tslS, def,
    external,requires, guarded, by},
  morecomment=[s]{\{-}{-\}},
  morecomment=[l]--,
  moredelim=[is][\emph]{'}{'},
  columns=fullflexible,
  flexiblecolumns=true,
   literate=
           {<}{$\langle$}1 {>}{$\rangle$}1
           {exists}{$\exists$}1
           {existsL}{$\existsl$}1
           {r0}{$\mathsf{r_0}$}2 {r1}{$\mathsf{r_1}$}2
           {r2}{$\mathsf{r_2}$}2 {r3}{$\mathsf{r_3}$}2
           {r4}{$\mathsf{r_4}$}2 {r5}{$\mathsf{r_5}$}2
           {r6}{$\mathsf{r_6}$}2 {r7}{$\mathsf{r_7}$}2
           {r8}{$\mathsf{r_8}$}2 {r9}{$\mathsf{r_9}$}2
           {rho5}{$\rho_5$}2
           {rho6}{$\rho_6$}2
           {rho7}{$\rho_7$}2
           {rho8}{$\rho_8$}2
           {0b}{\textbf{0} }2
           {^M2}{$^\mathsf{m_2}$}2
           {m2}{$\mathsf{m_2}$}2
           {^m}{$^\mathsf{m}$}1
           {m1}{$\mathsf{m_1}$}2
           {^L2}{$^\mathsf{l_2}$}2
           {l2}{$\mathsf{l_2}$}2
           {^l}{$^\mathsf{l}$}1
           {l1}{$\mathsf{l_1}$}2
           {forall}{$\forall$}1
           {<[}{$\ \stackrel{\text{def}}{=}\ $}3
           {]>}{}0
           {readonly}{\textbf{read-only}}9
           {taue}{$\tau_e$}2 {taum}{$\tau_m$}2
           {alpha}{$\alpha$}1 {beta}{$\beta$}1 {gamma}{$\gamma$}1
           {theta}{$\theta$}1 {lambda}{$\lambda$}1
           {delta}{$\delta$}1 {epsilon}{$\epsilon$}1 {tau}{$\tau$}1 
           {alphae}{$\alpha_e$}2
           {>^ro}{$\rangle^{\text{\textbf{ro}}}$}3
           {>^alpha}{$\rangle^\alpha$}2
           {>^beta}{$\rangle^\beta$}2
           {>^gamma}{$\rangle^\gamma$}2
           {>^lambda}{$\rangle^\lambda$}2
           {rec\ }{$\mu\ $}3
           {lock(lambda)}{$\lambda$}1
}

\lstdefinelanguage{miltypes}[]{mil}{
   literate=
           {<}{$\langle$}1 {>}{$\rangle$}1
           {exists}{$\exists$}1
           {existsL}{$\existsl$}1
           {r0}{$\mathsf{r_0}$}2 {r1}{$\mathsf{r_1}$}2
           {r2}{$\mathsf{r_2}$}2 {r3}{$\mathsf{r_3}$}2
           {r4}{$\mathsf{r_4}$}2 {r5}{$\mathsf{r_5}$}2
           {r6}{$\mathsf{r_6}$}2 {r7}{$\mathsf{r_7}$}2
           {r8}{$\mathsf{r_8}$}2 {r9}{$\mathsf{r_9}$}2
           {xrx}{$\register{}$}1
           {xrsx}{$\register s$}2
           {xrmx}{$\register m$}2
           {xrqx}{$\register q$}2
           {xrex}{$\register e$}2
           {xrlx}{$\register l$}2
           {xrvx}{$\register v$}2
           {xrcx}{$\register c$}2
           {xvx}{$v$}1
           {forall}{$\forall$}1 {mu}{$\mu$}1
           {<[}{$\ \stackrel{\text{def}}{=}\ $}3
           {]>}{}0
           {taue}{$\tau_e$}1 {taum}{$\tau_m$}1
           {readonly}{\textbf{read-only}}9
           {taue}{$\tau_e$}2 {taum}{$\tau_m$}2
           {alpha}{$\alpha$}1 {beta}{$\beta$}1 {gamma}{$\gamma$}1
           {theta}{$\theta$}1 {lambda}{$\lambda$}1
           {delta}{$\delta$}1 {epsilon}{$\epsilon$}1 {tau}{$\tau$}1 
           {alphae}{$\alpha_e$}2
           {>^ro}{$\rangle^{\text{\textbf{ro}}}$}3
           {>^alpha}{$\rangle^\alpha$}2
           {>^beta}{$\rangle^\beta$}2
           {>^gamma}{$\rangle^\gamma$}2
           {>^lambda}{$\rangle^\lambda$}2
           {rec\ }{$\mu\ $}3
           {lock(lambda)}{$\lambda$}1
           {def }{}0
           {=}{$\ \stackrel{\text{def}}{=}\ $}3
}

\newcommand{\pol}[1]{\vec{#1}}
\title{Type Inference for Deadlock Detection in a Multithreaded
  Polymorphic Typed Assembly Language}
\def\titlerunning{Type Inference for Deadlock Detection in MIL}

\author{Vasco T. Vasconcelos
\institute{LASIGE \& DI-FCUL,\\
   University of Lisbon, Portugal
   \email{vv@di.fc.ul.pt}} 
\and 
Francisco Martins 
\institute{LASIGE \& DI-FCUL,\\
   University of Lisbon, Portugal
   \email{fmartins@di.fc.ul.pt}}
\and 
Tiago Cogumbreiro
\institute{LASIGE \& DI-FCUL,\\
   University of Lisbon, Portugal
   \email{cogumbreiro@di.fc.ul.pt}}
}
\def\authorrunning{Vasconcelos, Martins, and Cogumbreiro}
\begin{document}
\maketitle

\begin{abstract}
  We previously developed a polymorphic type system and a type checker
  for a multithreaded lock-based polymorphic typed assembly language
  (MIL) that ensures that well-typed programs do not encounter race
  conditions. This paper extends such work by taking into
  consideration deadlocks.
  The extended type system verifies that locks are acquired in the
  proper order. Towards this end we require a language with
  annotations that specify the locking order.  Rather than asking the
  programmer (or the compiler's backend) to specifically annotate each
  newly introduced lock, we present an algorithm to infer the
  annotations.
  The result is a type checker whose input language is non-decorated as
  before, but that further checks that programs are exempt from deadlocks.
  %
\end{abstract}


\section{Introduction}

%
Type systems for lock-based race and deadlock static detection try to
contradict the idea put forward by some authors that ``the association
between locks and data is established mostly by
convention''~\cite{shavit:transactions-are-tomorrow-loads}.
Despite all the pathologies usually associated with locks (in the
aforementioned article and others), and specially at system's level,
locks are here to stay~\cite{cantrill.bonwick:real-world-concurrency}.


Deadlock detection should be addressed at the appropriate level
of abstraction, for, in general, compiled code that does not deadlock
allows us to conclude nothing of the source code.
Nevertheless, the problem remains valid at the assembly level and fits
quite nicely in the philosophy of typed assembly
languages~\cite{m.walker.c.g:sytemf-to-tal}.
By capturing a wider set of semantic properties, including the absence
of deadlocks, we improve compiler certification in systems where code
must be checked for safety before execution, in particular those with
untrusted or malicious components.


Our language targets a shared-memory machine featuring an array of
processors and a thread pool common to all
processors~\cite{cogumbreiro.martins.vasconcelos:compiling-pi-into-mtal,vasconcelos.martins:multithreaded-tal}. The
thread pool holds threads for which no processor is available, a
scheduler chooses a thread from this pool should a processor become
idle.  Threads voluntary release processors---our model fits in the
cooperative multi-threading category. For increased flexibility (and
unlike many other models,
including~\cite{flanagan.abadi:types-safe-locking}) we allow forking
threads that hold locks, hence we allow the suspension of processes
while in critical regions.
A prototype implementation can be found at
\href{http://gloss.di.fc.ul.pt/mil}{http://gloss.di.fc.ul.pt/mil}.

\begin{figure}[t]
\begin{lstlisting}
main () {
  f1,r3 := newLock; f3,r5 := newLock; f2,r4 := newLock -- 3 forks
  r1 := r3; r2 := r4; fork liftLeftFork[f1,f2]  -- 1st philosopher
  r1 := r4; r2 := r5; fork liftLeftFork[f2,f3]  -- 2nd philosopher
  r1 := r5; r2 := r3; fork liftLeftFork[f3,f1]  -- 3rd philosopher
  done
}
liftLeftFork forall[l,m].(r1:<l>^l, r2:<m>^m) {
  r3 := testSetLock r1
  if r3 = 0b jump liftRightFork[l,m]
  jump liftLeftFork[l,m]
}
liftRightFork forall[l,m].(r1:<l>^l, r2:<m>^m) requires {l} {
  r3 := testSetLock r2
  if r3 = 0b jump eat[l,m]
  jump liftRightFork[l,m]
}
eat forall[l,m].(r1:<l>^l, r2:<m>^m) requires {l,m} {
  -- eat
  unlock r1 -- lay down the left fork
  unlock r2 -- lay down the right fork
  -- think
  jump liftLeftFork[l,m]
}
\end{lstlisting}
  \caption{The dining philosophers written in MIL}
  \label{fig:philosophers}
\end{figure}

The code in Figure~\ref{fig:philosophers} presents a typical example
of a potential deadlock comprising a cycle of threads where each
thread requests a lock hold by the next thread.  Imagine the code
running on a two-processors machine: after \lstinline|main| completes
its execution, each philosopher embarks on a busy-waiting loop, only
that two of them will be running in processors, while the third is
(and will indefinitely remain) in the run-pool.
Situations of deadlocks comprising suspended code are known to be
difficult to deal
with~\cite{kontothanassis.etal:scheduler-conscious-sync}.
Our notion of deadlocked state takes into account running and
suspended threads.

Another source of difficulties in characterizing deadlock states
derives from the low-level nature of our language that decouples the
action of lock acquisition from that of entering a critical section,
and that features non-blocking instructions only.
As such the meaning of ``entering a critical section'' cannot be of a
syntactic nature.

A characteristic of our machine is the syntactic dissociation of the
test-and-set-lock and the jump-to-critical operations, for which we
provide two distinct instructions, as found in conventional
instruction sets. Furthermore, there is no syntactic distinction
between a conventional conditional jump and a (conditional)
jump-to-critical instruction, and the test-set-lock and
jump-to-critical instructions can be separated by arbitrary assembly
code.  As far as the type system goes, the thread holds the lock only
after the conditional jump, even though at runtime it may have been
obtained long before.

The main contribuitons of this paper are:
\begin{itemize}
\item A type system for deadlock elimination. We devise a type system
  that establishes a strict partial order on lock acquisition, hence
  enforcing that well typed MIL programs do not
  deadlock---Theorem~\ref{thm:ts};
\item An algorithm for automatic program annotation. In order to check
  the absence of deadlock, MIL programs must be annotated to reflect
  the order by which locks must be acquired.
  Annotating large assembly programs, either manually or as the result
  of a compilation process, is not plausible.
  We present an algorithm that takes a plain MIL program and produces
  an annotated program together with a collection of constraints over
  lock sets that are passed to a constraint solver. In case the
  constraints are solvable the annotated program is
  typeble---Theorem~\ref{thm:soundness}---hence free from deadlocks.
\end{itemize}

The outline of this paper is as follows. The next section introduces
the syntax of programs and machine states, together with the running
example. Then Section~\ref{sec:semantics} presents the operational
semantics and the notion of deadlocked
states. Section~\ref{sec:typing} describes the type system and the
first main result, typable states do not
deadlock. Section~\ref{sec:type-inference} introduces the annotation
algorithm and the second main result, the correctness of the algorithm
with respect to the type system. Finally, Section~\ref{sec:conclusion}
describes related work and concludes the paper.


\begin{myfigure}
\begin{alignat*}2
  & \textit{registers} &
  r \grmeq& \register 1 \grmor \dots \grmor \register\numreg
  \\
  & \textit{lock values} &
  b \grmeq& \lockbit {0} \grmor \lockbit 1  \grmor \lockbit {0}^\lambda
  \\
  & \textit{values} &
  v \grmeq& r \grmor n
  \grmor b
  \grmor l 
  \grmor \instantiation\lambda
  \grmor \uninit
  \\
  & \textit{instructions} &
  \iota \grmeq&
  \\
  & \quad\textit{control flow} &&
    \movei \grmor
    \aopi \grmor
    \cjumpi \grmor
    \forki
  \\
   & \quad\textit{memory} &&
   \malloci \grmor
     \loadi \grmor
     \storei \grmor
   \\
  & \quad\textit{locking} &&
    \newlocki \grmor
    \tsli \grmor
    \unlocki
  \\
  & \textit{inst.\ sequences}\quad &
  I \grmeq&
  \iota ; I \grmor
  \jumpi \grmor
  \donei\\
    & \textit{types} &
    \tau \grmeq&
    \integer \grmor
    \lambda \grmor
    \tupletypet \grmor
    \codetypet \grmor
    \forallt
    \\
    & \textit{register file types}\quad &
    \Gamma \grmeq& \register 1\colon \tau_1,\dots,\register n\colon\tau_n
    \\
    & \textit{permissions} &
    \Lambda \grmeq& \lambda_1,\dots,\lambda_n
    \\
 & \textit{heaps} &
  H \grmeq& \{l_1 \Mapsto h_1, \dots, l_n \Mapsto h_n\}
  \\
  & \textit{heap values}\quad &
  h \grmeq& \heaptuple{v_1\dots v_n}{\lambda} \grmor \heapcodev
  \\
  & \textit{thread pool} &
  T \grmeq& \{\poolitem{l_1[\vec\lambda_1]}{R_1},\dots,\poolitem{l_n[\vec\lambda_n]}{R_n}\}
  \\
  & \textit{register files} &
  R \grmeq& \{\register 1 \Mapsto v_1, \dots, \register\numreg \Mapsto v_\numreg\}
  \\
  & \textit{processors array}\qquad &
  P \grmeq& \{1 \Mapsto p_1, \dots, \numthr \Mapsto p_{\numthr}\}
  \\
  & \textit{processor}\qquad &
  p \grmeq& \processor R \Lambda I
  \\
  & \textit{states} &
  S \grmeq& \program H T P \grmor \halt
\end{alignat*}
\caption{Syntax.}
\label{fig:syntax-instructions}
\end{myfigure}



\section{Syntax}
\label{sec:syntax}

The syntax of our language is generated by the grammar in
Figure~\ref{fig:syntax-instructions}. %
We rely on two mutually disjoint sets for \emph{heap labels}, ranged
over by~$l$, and for \emph{singleton lock types}, ranged over by
$\lambda$. Letter $n$ ranges over integer values.

Values $v$ comprise registers $r$, integer values $n$, lock values
$b$, labels $l$, type application $v[\lambda]$, and uninitialised
values~$\uninit$.
Lock value $\lockbit 0$ represents an open lock, whereas lock value
$\lockbit 1$ denotes a closed lock; the~$\lambda$ annotation in
$\lockbit 0^\lambda$ allows to determine the lock guarding the
critical section a processor is trying to enter and will be useful
when defining deadlocked states. Lock values are runtime entities,
they need to be distinct from conventional integer values for typing
purposes only.
Labels are used as heap addresses.
Uninitialised values represent meaningless data of a certain
type.

Most of the machine instructions $\iota$ presented in
Figure~\ref{fig:syntax-instructions} are standard in assembly
languages.
Distinct in MIL are the instructions for creating new
threads---$\fork$ places in the run queue a new thread waiting for
execution---, for allocating memory---$\malloc \mallocarg {\tau_1,
  \dots, \tau_n} \lambda$ allocates a tuple in the heap protected by
lock $\lambda $ and comprising $n$ cells each of which containing an
uninitialised value of type $\tau_i$---, and for manipulating locks. In
this last group one finds $\newlock$ to create a lock in the heap and
store its address in register $r$ ($\lambda$ describes the singleton
lock type associated to the new lock, further described below), $\tsl$
to acquire a lock, and $\unlock$ to release a lock.

Instructions are organised in sequences~$I$, ending in $\jump$ or in
$\done$. 
Instruction \done terminates a thread, voluntarily releasing the core,
giving rise to a cooperative multi-threading model of computation.

Types $\tau$ include the integer type $\integer$, the singleton lock
type~$\lambda$, the tuple type $\tupletypet$ describing a tuple in the
heap protected by lock~$\lambda$, and the code type $\codetypea$
representing a code block abstracted on singleton lock types
$\pol\lambda$, expecting registers of the types in $\Gamma$ and
requiring locks as in $\Lambda$. Each universal variable is bound by
two sets of singleton lock types $\Lambda$, used for deadlock
prevention, as described below.
%
%
For simplicity we allow
polymorphism over singleton lock types only; for abstraction over
arbitrary types
see~\cite{cogumbreiro.martins.vasconcelos:compiling-pi-into-mtal}.

The \emph{abstract machine} is parametric on the number of
available processors $\numthr$, and on the number of registers per
processor~$\numreg$.
An abstract machine can be in two possible states $S$: halted or
running. A running machine comprises a heap $H$, a thread pool $T$,
and an array of processors $P$ of fixed length~$\numthr$.
Heaps are maps from labels $l$ into \emph{heap values} $h$ that may be
either data tuples or code blocks.
\emph{Tuples} $\heaptuple{v_1, \dots, v_n}\lambda$ are vectors of
mutable values $v_i$ protected by some lock $\lambda$.
Code blocks $\heapcode \codetypea I$ comprise a signature (a code type) and
an instruction sequence~$I$, to be executed by a processor.
A thread pool~$T$ is a multiset of pairs
$\poolitem{l[\vec\lambda]}{R}$, each of which contains the address (a
label) of a code block in the heap, a sequence of singleton lock types
to act as arguments to the forall type of the code block, and a
register file.
A processor array $P$ contains~$\numthr$ processors, each of which is
composed of a register file $R$ mapping the processor's registers to
values, a set of locks $\Lambda$ (the locks held by the thread running
at the processor, often call the thread's \emph{permission}), and a
sequence of instructions $I$ (the instructions that remain to
execute).

\paragraph{Lock order annotations}
Deadlocks are usually prevented by imposing a strict partial order on
locks, and by respecting this order when acquiring
locks~\cite{birrell:programming-with-threads,coffman.etal:system-deadlocks,flanagan.abadi:types-safe-locking}.
The syntax in Figure~\ref{fig:syntax-instructions} introduces
annotations that specify the locking order.
When creating a new lock, we declare the order between the newly
introduced singleton lock type and the locks known to the program.  We
use the notation $\lambda\colon(\Lambda_1,\Lambda_2)$ to mean that
lock type $\lambda$ is greater than all lock types in set $\Lambda_1$
and smaller than each lock type in set $\Lambda_2$.
The annotated syntax differs from the original
syntax (\cite{cogumbreiro.martins.vasconcelos:compiling-pi-into-mtal,vasconcelos.martins:multithreaded-tal})
in two places:
at lock creation $\lambda\colon(\Lambda,\Lambda),r \assign \newlock$;
and in universal types $\forallt$, where we explicitly specify the
lock order on newly introduced singleton lock types.

\paragraph{Example}
Figure~\ref{fig:philosophers} shows an example of a
\emph{non-annotated} program. Annotating such a program requires
describing the order for each lock introduced in code block
\lstinline|main|, say,
\begin{lstlisting}
f1::({},{}), r3 := newLock
f3::({f1},{}), r5 := newLock
f2::({f1},{f3}), r4 := newLock  
\end{lstlisting}
and at the types for the three code blocks below.
\begin{lstlisting}
liftLeftFork forall[l::({},{})].forall [m::({l},{})].(r1:<l>^l, r2:<m>^m)
liftRightFork forall[l::({},{})].forall[m::({l},{})].(r1:<l>^l, r2:<m>^m) requires {l}
eat forall[l::({},{})].forall[m::({l},{})].(r1:<l>^l, r2:<m>^m) requires {l,m}
\end{lstlisting}

%
Notice that abstracting one lock at a time, as in the types just shown,
precludes declaring code blocks with non-strict partial orders on
locks, such as $\forall[l\colon(\emptyset,\{m\}),
m\colon(\{l\},\emptyset)].\tau$, which cannot be fulfilled by any
conceivable sequence of instructions.



\begin{myfigure}
  \begin{gather*}
    \tag\Rhalt
    \frac{
      \forall i. P(i) = \processor{\_\ }{\_\ }{\done}
    }{
      \program \_ \emptyset P \rightarrow \halt
    }
    \\[\rulespc]
    \tag\Rschedule
    \frac{
      P(i) = \processor{\_\ }{\_\ }{\done} \qquad 
      H(l) = \heapcode {\codetypeforall {\pol \lambda\colon(\_,\_)} {\_} \Lambda} I
    }{
      \program{H}{T\uplus\{\poolitem{l[\pol\lambda']}{R}\}}{P}
      \rightarrow 
      \program H T {P\update{i}{\processor{R}{\Lambda}{I}\subs{\pol\lambda'}{\pol\lambda}}}
    }
    \\[\rulespc]
    \tag\Rfork
    \frac{
      P(i) = \processor R {\Lambda\uplus\Lambda'} {(\forki;I)}
      \qquad
      \eval(v) = l[\pol \lambda]
      \qquad    
      H(l) = \heapcode {\codetypeforall {\_} {\_} {\Lambda'}} {\_}
    }{
      \program{H}{T}{P} \rightarrow \program{H}{T\cup\{\poolitem{l[\pol \lambda]}{R}\}}
      {P\update i {\processor R {\Lambda} {I}}}
    }
    \\[\rulespc]
 \tag\Rnewlock
  \frac{
    P(i) = \processor R \Lambda {(\lambda\colon(\_, \_),r \assign \newlock;I)} \quad
    l \not\in\dom(H) \quad
    \lambda'\text{ fresh}
    }{
    \program H T P \rightarrow
    \program {H\update{l}{\heaptuple{\lockbit 0}{\lambda'}}}
             {T} 
             {P\update i {\processor{R\update rl}{\Lambda}{I\subs{\lambda'}\lambda}}}
  }
  \\[\rulespc]
  \tag\RtslZ
  \frac{
    P(i) = \processor R \Lambda {(\tsli;I)} \quad
    \eval(v) = l \quad 
    H(l) = \heaptuple{\lockbit 0}{\lambda}
  }{
    \program H T P \rightarrow
    \program {H\update{l}{\heaptuple{\lockbit 1}\lambda}}
             {T}
             {P\update i {\processor{R\update r{\lockbit 0 ^\lambda}}{\Lambda \uplus \{\lambda\}}{I}}}
  }
  \\[\rulespc]
  \tag\RtslO
  \frac{
    P(i) = \processor R \Lambda {(\tsli;I)} \quad
    H(\eval(v)) = \heaptuple{\lockbit 1}{\lambda} \quad
    \lambda \not\in \Lambda
    }{
    \program H T P \rightarrow
    \program H T {P\update i {\processor{R\update r{\lockbit 1}}{\Lambda}{I}}}
  }
  \\[\rulespc]
  \tag\Runlock
  \frac{
    P(i) = \processor R {\Lambda\uplus\{\lambda\}} {(\unlocki;I)} \quad
    \eval(v) = l \quad 
    H(l) = \heaptuple{\_}{\lambda}
    }{
    \program H T P \rightarrow 
    \program{H\update l{\heaptuple{\lockbit 0}\lambda}}{T}{
      P\update i {\processor R\Lambda I}}
  }
 \end{gather*}
  \caption{Operational semantics (thread pool and locks).}
  \label{fig:reduction-runqueue}
\end{myfigure}



\begin{myfigure}
\begin{gather*}
  \tag\Rmalloc
  \frac{
    P(i) = \processor R \Lambda {(\malloci;I)}
    \qquad 
    l \notin \dom(H)
  }{
    \program H T P \rightarrow 
    \program {H \update{l}{\heaptuple{?\pol \tau}\lambda}}
             {T}
             {P\update i {\processor{R\update r l}
                 {\Lambda}{I}}}
  }
  \\[\rulespc]
  \tag\Rload
  \frac{
    P(i) = \processor R \Lambda {(\loadi;I)} \qquad
    H(\eval(v)) = \heaptuple{v_1..v_n..v_{n+m}}\lambda \qquad
    \lambda \in \Lambda
    }{
    \program H T P \rightarrow
    \program H T {P\update i {\processor{R\update r{v_n}}{\Lambda}{I}}}
  }
  \\[\rulespc]
  \tag\Rstore
  \frac{
    P(i) = \processor R \Lambda {(\storei;I)} \qquad
    R(r) = l \qquad
    H(l) = \heaptuple{v_1..v_n..v_{n+m}}{\lambda} \qquad
    \lambda \in \Lambda
  }{
    \program H T P \rightarrow 
    \program{H\update l{\heaptuple{v_1..\eval(v)..v_{n+m}}\lambda}}{T}{
      P\update i {\processor R \Lambda I}}
  }
  \\[\rulespc]
  \tag\Rjump
  \frac{
    P(i) = \processor R \Lambda \jumpi
    \qquad
    \eval(v) = l[\pol\lambda]
    \qquad
    H(l) = \heapcode{\forall[\pol\lambda'\colon(\_,\_)].\_}{I}
  }{
    \program H T P \rightarrow 
    \program H T {P\update i {\processor R \Lambda
        {I\subs{\pol\lambda}{\pol\lambda'}}}}
  }
  \\[\rulespc]
  \tag\Rmove
  \frac{
    P(i) = \processor R \Lambda {(\movei;I)}
  }{
    \program H T P \rightarrow 
    \program H T {P\update i {\processor {R\update{r}{\eval(v)}} \Lambda I}}
    }
  \\[\rulespc]
  \tag\Rarith
  \frac{
    P(i) = \processor R \Lambda {(\aopi[r'];I)}
  }{
    \program H T P \rightarrow
    \program H T {P\update i {\processor{R\update{r}{R(r') + \eval(v)}}{\Lambda}{I}}}
  }
  \\[\rulespc]
  \tag\RbranchT
  \frac{
    P(i) = \processor R \Lambda {(\cjump{r}{v}{v'};\_)}
    \qquad
    R(r) = v
    \qquad 
    \eval(v') = l[\pol\lambda]
    \qquad
    H(l) = \heapcode{\forall[\pol\lambda'\colon(\_,\_)].\_}{I}
  }{
    \program H T P \rightarrow
    \program H T {P\update i {\processor R \Lambda {I\subs{\pol\lambda}{\pol\lambda'}}}} 
  }
  \\[\rulespc]
  \tag\RbranchF
  \frac{
    P(i) = \processor R \Lambda {(\cjump{r}{v}{\_};I)}
    \qquad
    R(r)\not= v 
    }{
    \program H T P \rightarrow 
    \program H T {P\update i {\processor R \Lambda I}} 
  }
\end{gather*}
\caption{Operational semantics (memory and control flow).}
\label{fig:reduction-memory}
\end{myfigure}


\section{Operational Semantics and Deadlocked States}
\label{sec:semantics}

The operational semantics is defined in
Figures~\ref{fig:reduction-runqueue} and~\ref{fig:reduction-memory}.
The scheduling model of our machine is described by the first three
rules in Figure~\ref{fig:reduction-runqueue}.
The machine halts when all processors are idle and the thread pool is
empty (rule $\Rhalt$).
An idle processor (a processor that executes instruction $\done$)
picks up an arbitrary thread from the thread pool and activates it
(rule $\Rschedule$); the argument locks $\pol\lambda'$ replace the
parameters $\pol\lambda$ in the code for the processor.
For a $\fork$ instruction, the machine creates a ``closure'' by
putting together the code label plus its arguments, $l[\pol \lambda]$,
and a copy of the registers, $R$, and by placing it in the thread
pool.  The thread permission is partitioned in two: one part
($\Lambda$) stays with the thread, the other ($\Lambda'$) goes with
the newly created thread, as required by the type of its code.

Some rules rely on the evaluation function~$\eval$ that looks for
values in registers and in value application.
\begin{equation*}
  \eval(v) = 
  \begin{cases}
    R(v) & \text{if } v \text{ is a register}\\
    \eval(v')[\lambda] & \text{if } v \text{ is } v'[\lambda]\\
    v & \text{otherwise}
  \end{cases}
\end{equation*}

In our model the heap tuple~$\tupletype{\lockbit 0}{\lambda}$
represents an open lock, whereas~$\tupletype{\lockbit 1}{\lambda}$
represents a closed lock.  A lock is an uni-dimensional tuple holding
a \emph{lock value} because the machine provides for tuple allocation
only; lock $\lambda$ is used for type safety purposes, just like all
other singleton lock types.
Instruction~$\newlock$ creates a new open lock in the heap and places
a reference $l$ to it in register~$r$.
Instruction $\tsl$ loads the contents of the lock tuple into
register~$r$ and sets the heap value to~$\tupletype{\lockbit
  1}{\lambda}$; it also makes sure that the lock is not in the thread's
permission (rules~$\RtslZ$ and~$\RtslO$).
Further, applying the instruction to an unlocked lock adds
lock~$\lambda$ to the permission of the processor (rule~$\RtslZ$).
Locks are waved using instruction $\unlock$, as long as the thread
holds the lock (rule~$\Runlock$).


Rules related to memory manipulation are described in
Figure~\ref{fig:reduction-memory}.
Rule~$\Rmalloc$ creates an heap-allocated $\lambda$-protected
uninitialised tuple and moves its address to register~$r$.
To store values in, and load from, a tuple we require that the lock
that guards the tuple is among the processor's permission.
%
%
In rules \RbranchT and \RbranchF, we ignore the lock annotation on
lock values, so that $\lockbit 0^\lambda$ is considered equal to
$\lockbit{0}$.  The remaining rules are standard
(cf~\cite{m.walker.c.g:sytemf-to-tal}).


\paragraph{Deadlocked States}

%
The difficulty in characterising deadlock states stems from the fact
that processors never block and that threads may become (voluntary)
suspended while in critical a region. We aim at capturing conventional
techniques for acquiring locks, namely busy-waiting and
sleep-lock~\cite{vasconcelos.martins:multithreaded-tal}.
Towards this end, we need to restrict reduction of a given state $S$
to that of a single processor in order to control the progress of a
single core: let relation~$S \rightarrow_i S'$ denote a reduction step
on processor $i$ excluding rules \Rhalt, \Rschedule and \Runlock.

\begin{defi}[Deadlocked states]
  Let $S$ be the state $\program HTP$.
  \begin{itemize}
 \item A processor $\processor R\Lambda I$ holds lock $\lambda$ when
    $\lambda\in\Lambda$; a suspend thread
    $\poolitem{l[\vec\lambda']}{R}$ holds lock $\lambda$ when $H(l) =
    \heapcode {\codetypeforall {\pol \lambda\colon(\_,\_)} {\_}
      \Lambda} \_$ and $\lambda\in
    \Lambda\subs{\pol\lambda'}{\pol\lambda}$;
  \item A \emph{processor $p$ in $P$ immediately tries to enter a
      critical section guarded by lock $\lambda$} if $p$ is of the
    form $\processor{R}{\_}{(\cjumpli;\_)}$ and $R(r) = \lockbit
    0^\lambda$;
  \item For busy waiting, a \emph{thread in processor $p_i$ is
      trying to enter a critical region guarded by $\lambda$} if $S
    \rightarrow^*_i S'$ and processor $p_i$ in state~$S'$ immediately
    tries to enter a critical section guarded by~$\lambda$;
  \item For sleep-lock, a \emph{thread
      $\poolitem{l[\pol\lambda']}{R}$ in thread pool $T$ is trying to enter a
      critical region guarded by $\lambda$} if $ H(l) = \heapcode
    {\codetypeforall {\pol\lambda\colon\_} {\_} \Lambda} I$, and
    the thread in processor $p_1$ of state $S + P\{1\colon
    \processor{R}{\Lambda}{I}\subs{\pol\lambda'}{\pol\lambda}\}$ is
    trying to enter a critical region guarded by $\lambda$;
  \item A state $S$ is \emph{deadlocked} if there exist locks
    $\lambda_0,\dots,\lambda_n$, with $\lambda_0=\lambda_n$, and
    indices~$d_0,\dots,d_{n-1}$ ($n>0$) such that for each $0\le i<n$,
    either processor $p_{d_i}$ or suspended thread $t_{d_i}$ holds lock
    $\lambda_i$ and is trying to enter a critical region guarded by
    $\lambda_{i+1}$.
  \end{itemize}
\end{defi}

Notice that $d_i\neq d_j$ does not imply $p_{d_i} \neq p_{d_j}$ and
similarly for threads in the thread pool, so that a state deadlocked
on locks $\lambda_0,\dots,\lambda_n$ may involve less than $n$
threads.  We have excluded the \Runlock rule from the $\rightarrow_i$
reduction relation, yet releasing a lock is not necessarily an
indication that the thread is leaving a deadlocked state, for the
released lock may not be involved in the deadlock; a more general
definition of deadlocked state would take this fact into account.



\begin{myfigure}
  \begin{gather*}
   \frac{
      \Psi \vdash \lambda\colon (\Lambda_1,\_)
      \qquad
      \lambda_1 \in \Lambda_1
    }{
      \Psi \vdash \lambda_1 \prec \lambda
    }
    \qquad
    \frac{
      \Psi \vdash \lambda\colon (\_,\Lambda_2)
      \qquad
      \lambda_2 \in \Lambda_2
    }{
      \Psi \vdash \lambda \prec \lambda_2
    }
    \qquad
    \frac{
      \Psi \vdash \lambda_1 \prec \lambda_2
      \qquad
      \Psi \vdash \lambda_2 \prec \lambda_3
    }{
      \Psi \vdash \lambda_1 \prec \lambda_3
    }
    \\
    \frac{
      \Psi\vdash\lambda_1\prec\lambda \quad\cdots\quad \Psi\vdash\lambda_n\prec\lambda
    }{
      \Psi\vdash\{\lambda_1,\dots,\lambda_n\} \prec \lambda
    }
    \qquad
    \frac{
      \Psi\vdash\lambda\prec\lambda_1 \quad\cdots\quad \Psi\vdash\lambda\prec\lambda_n
    }{
      \Psi\vdash \lambda \prec \{\lambda_1,\dots,\lambda_n\}
    }
  \end{gather*}
  \caption{Less-than relation on locks and permissions}
  \label{fig:less-than}
\end{myfigure}

\begin{myfigure}
  \begin{gather*}
    \tag{\Ttype,\SregFile}
    \frac{
      \ftv(\tau) \subseteq \dom(\Psi)
    }{
      \Psi\vdash\tau
    }
    \qquad
    \frac{
      \Psi \vdash \tau_i
    }{
      \Psi
      \vdash
      \register 1\colon\tau_1,\dots,\register {n+m}\colon\tau_{n+m}
      \subtype
      \register 1\colon\tau_1,\dots,\register n\colon\tau_n
    }
    \\[\rulespc]
    \tag{\Tlabel,\Treg,\Tint,\Tlock,\Tuninit}
    \frac{
      \Psi \vdash \tau
    }{
      \Psi, l \colon \tau;\Gamma \vdash l\colon \tau
    }
    \qquad
    \frac{
      \Psi \vdash \tau
    }{
      \Psi; \Gamma_1,r_i\colon\tau,\Gamma_2 \vdash r_i\colon \tau
    }
    \qquad
    \Psi;\Gamma \vdash n\colon \integer
    \qquad
    \Psi; \Gamma \vdash \lockbit 0,\lockbit 1,\lockbit 0^\lambda \colon \lambda
   \qquad
    \Psi; \Gamma \vdash \uninit \colon \tau
    \\[\rulespc]
    \tag{\Tvalapp}
    \frac{
      \Psi\vdash\lambda'
      \qquad
      \Psi; \Gamma \vdash v \colon
      \forall[\lambda\colon(\Lambda_1,\Lambda_2)]\tau
      \qquad
      \Psi \vdash \Lambda_1 \prec \lambda' \prec \Lambda_2
    }{
      \Psi;\Gamma \vdash \instantiation{\lambda'} \colon 
      \tau
      \subs{\lambda'}{\lambda}
    }
  \end{gather*}
  \caption{Rules for values
    \myfbox{$\Psi;\Gamma \vdash v\colon\tau$}, for subtyping
    \myfbox{$\Psi\vdash \Gamma \subtype \Gamma$}, and for types
    \myfbox{$\Psi \vdash \tau$}.}
  \label{fig:static-values}
\end{myfigure}


\begin{myfigure}
  \begin{gather*}
    \tag{\Tdone}
    \Psi;\Gamma;\emptyset \vdash \donei
    \\[\rulespc]
    \tag{\Tfork}
    \frac{
      \Psi;\Gamma \vdash v\colon \codetype{\Gamma'}{\Lambda}
      \qquad
      \Psi;\Gamma;\Lambda'\vdash I
      \qquad
      \Psi \vdash \Gamma \subtype \Gamma' 
    }{
      \Psi;\Gamma;\Lambda\uplus\Lambda' \vdash \forki;I
    }
    \\[\rulespc]
    \tag{\Tnewlock}
    \frac{
      \Psi,\lambda\colon(\Lambda_1, \Lambda_2);\Gamma\update{r}{\tupletype{\locktype}\lambda};
      \Lambda \vdash I \qquad
      \lambda \not \in \Psi, \Gamma, \Lambda
    }{
      \Psi;\Gamma;\Lambda \vdash \lambda\colon(\Lambda_1, \Lambda_2),r \assign \newlock; I
    }
    \\[\rulespc]
    \tag{\Ttsl}
    \frac{
      \Psi;\Gamma \vdash v\colon \tupletype\locktype\lambda
      \qquad
      \Psi;\Gamma\update{r}{\locktype};\Lambda \vdash I
      \qquad
      \lambda \not \in \Lambda
    }{
      \Psi;\Gamma;\Lambda \vdash \tsli; I
    }
    \\[\rulespc]
    \tag{\Tunlock}
    \frac{
      \Psi;\Gamma \vdash v\colon\tupletype\locktype\lambda
      \qquad
      \Psi;\Gamma;\Lambda \vdash I
    }{
      \Psi;\Gamma;\Lambda\uplus\{\lambda\} \vdash \unlocki; I
    }
    \\[\rulespc]
    \tag{\TbranchC}
    \frac{
      \Psi;\Gamma \vdash r\colon \locktype
      \qquad
      \Psi;\Gamma \vdash v\colon\codetype{\Gamma'}{\Lambda\uplus\{\alpha\}}
      \qquad
      \Psi;\Gamma;\Lambda \vdash I
      \qquad
      \Psi \vdash \Gamma \subtype \Gamma'
      \qquad
      \Psi \vdash \Lambda \prec \lambda
    }{
      \Psi;\Gamma;\Lambda \vdash \cjumpli; I
    }
  \end{gather*}
  \caption{Typing rules for instructions (thread pool and locks)
    \myfbox{$\Psi;\Gamma;\Lambda \vdash I$}.}
  \label{fig:static-instructions-pool}
\end{myfigure}


\begin{myfigure}
  \begin{gather*}
    \tag{\Tmalloc}
    \frac{
      \Psi;
      \Gamma\update r {\tupletype{\pol \tau}\lambda};\Lambda \vdash I
      \qquad
      \tau_i \neq \locktype[\lambda]
      \qquad
      \lambda \in \Lambda
    }{
      \Psi;\Gamma;\Lambda \vdash \malloci; I
    }
    \\[\rulespc]
    \tag{\Tload}
    \frac{
      \Psi;\Gamma\vdash v\colon
      \tupletype{\tau_1..\tau_{n+m}}{\lambda}
      \qquad
      \Psi;\Gamma\update r{\tau_n};\Lambda \vdash I
      \qquad
      \tau_n \neq \lambda'
      \qquad
      \lambda\in\Lambda
    }{
      \Psi;\Gamma;\Lambda \vdash \loadi;I
    }
    \\[\rulespc]
     \tag{\Tstore}
     \frac{
       \Psi;\Gamma \vdash v\colon\tau_n
       \quad\;\;
       \Psi;\Gamma \vdash r\colon
       \tupletype{\tau_1..\tau_{n+m}}{\lambda}
       \quad\;\;
       \Psi;\Gamma\update{r}{\tupletype{\tau_1..\tau_{n+m}}{\lambda}};\Lambda \vdash I
       \quad\;\;
       \tau_n \neq \lambda'
       \quad\;\;
       \lambda\in\Lambda
     }{
       \Psi;\Gamma;\Lambda \vdash \storei; I 
     }
    \\[\smallrulespc]
    \tag{\Tmove}
    \frac{
      \Psi; \Gamma \vdash v\colon \tau
      \qquad
      \Psi;\Gamma\update r\tau;\Lambda \vdash I
    }{
      \Psi;\Gamma;\Lambda \vdash \movei;I
    }
    \\[\rulespc]
    \tag{\Tarith}
    \frac{
      \Psi; \Gamma \vdash r'\colon \integer
      \qquad
      \Psi;\Gamma\vdash v\colon \integer
      \qquad
      \Psi;\Gamma\update r\integer;\Lambda \vdash I
    }{
      \Psi;\Gamma;\Lambda \vdash \aopi[r'];I
    }
    \\[\rulespc]
    \tag{\Tbranch}
    \frac{
      \Psi; \Gamma \vdash r\colon \integer
      \qquad
      \Psi; \Gamma \vdash v\colon \integer
      \qquad
      \Psi; \Gamma \vdash v\colon\codetype{\Gamma}{\Lambda}
      \qquad
      \Psi;\Gamma;\Lambda \vdash I
    }{
      \Psi;\Gamma;\Lambda \vdash \cjumpi[v];I
    }
    \\[\rulespc]
    \tag{\Tjump}
    \frac{
      \Psi;\Gamma \vdash v\colon\codetype{\Gamma'}{\Lambda}
      \qquad
      \Psi \vdash \Gamma \subtype \Gamma'
    }{
      \Psi;\Gamma;\Lambda \vdash \jumpi
    }
  \end{gather*}
  \caption{Typing rules for instructions (memory and control flow)
    \myfbox{$\Psi;\Gamma;\Lambda \vdash I$}.}
  \label{fig:static-instructions-memory}
\end{myfigure}


\begin{myfigure}
  \begin{gather*}
    \tag{reg file, \fbox{$\Psi \vdash R \colon \Gamma$}}
    \frac{
      \forall i.
      \Psi \vdash \Gamma(\register i)
      \qquad
      \Psi;\emptyset \vdash R(\register i)\colon \Gamma(\register i)
   }{
      \Psi\vdash R\colon\Gamma
    }
    \\[\rulespc]
    \tag{processors, \fbox{$\Psi \vdash P$}}
    \frac{
      \forall i.\Psi \vdash P(i)
    }{
      \Psi \vdash P
    }
    \qquad
    \frac{
      \Psi \vdash R\colon\Gamma
      \qquad
      \Psi;\Gamma';\Lambda \vdash I
      \qquad
      \Psi \vdash \Gamma \subtype \Gamma'
    }{
      \Psi \vdash \processor{R}{\Lambda}{I}
    }
    \\[\rulespc]
    \tag{thread pool, \fbox{$\Psi \vdash T$}}
    \frac{
      \forall i.
      \Psi \vdash t_i
    }{
      \Psi \vdash \{t_1,\dots,t_n\}
    }
    \qquad
    \frac{
      \Psi;\emptyset\vdash v\colon 
      \codetype {\Gamma'} \_
      \qquad 
      \Psi\vdash R\colon \Gamma
      \qquad
      \Psi \vdash \Gamma \subtype \Gamma'
    }{
      \Psi \vdash \poolitem{v}{R}
    }
    \\[\rulespc]
    \tag{heap value, \fbox{$\Psi \vdash h\colon\tau$}}  
    \frac{
      \tau = \codetypeforall{\pol\lambda\colon(\pol \Lambda_1, \pol \Lambda_2)}{\Gamma}{\Lambda}
      \qquad
      \Psi, \pol\lambda\colon(\pol \Lambda_1, \pol \Lambda_2);\Gamma;\Lambda \vdash I
    }{
      \Psi \vdash \tau \{I\} \colon \tau
    }
   \qquad
   \frac{
      \forall i.\Psi;\emptyset \vdash v_i \colon \tau_i
   }{
      \Psi \vdash \heaptuple{\pol v}{\lambda} \colon 
      \tupletype{\pol \tau}\lambda
    }
    \\[\rulespc]
    \tag{heap, \fbox{$\Psi \vdash H$}}
    \frac{
      \forall l. \Psi \vdash H(l) \colon \Psi(l)
    }{
      \Psi \vdash H
    }
    \\[\rulespc]
    \tag{state, \fbox{$\Psi \vdash S$}}
    \Psi \vdash \halt
    \qquad
    \frac{
      \Psi \vdash H \qquad \Psi \vdash T \qquad \Psi \vdash P
    }{
      \Psi \vdash \program HTP
    }
  \end{gather*}
  \caption{Typing rules for machine states.}
  \label{fig:static-machine}
\end{myfigure}


\section{A Type System for Deadlock Prevention}
\label{sec:typing}

\myparagraph{Type System}
Typing environments $\Psi$ map heap addresses $l$ to types $\tau$, and
singleton lock types $\lambda$ to lock kinds
$(\Lambda_1,\Lambda_2)$. An entry $\lambda\colon(\Lambda_1,\Lambda_2)$
in $\Psi$ means that $\lambda$ is larger than all lock types in
$\Lambda_1$ and smaller than any lock type in $\Lambda_2$, a notion
captured by relation $\prec$ described in Figure~\ref{fig:less-than}.
Instructions are also checked against a register file type $\Gamma$
holding the current types of the registers, and a set $\Lambda$ of
lock variables: the \emph{permission} of (the processor executing) the
code block.
The type system is presented in Figures~\ref{fig:static-values}
to~\ref{fig:static-machine}.


Typing rules for values are illustrated in
Figure~\ref{fig:static-values}.
Rule $\Ttype$ makes sure types are \emph{well-formed}, that all free
singleton lock types (or free type variables, $\ftv$) in a type are
bound in the typing environment.
A formula $\Gamma \subtype \Gamma'$ allows ``forgetting'' registers in
the register file type, and is particularly useful in jump
instructions where we want the type of the target code block to be
more general (ask for less registers) than those active in the current
code~\cite{m.walker.c.g:sytemf-to-tal}.
The rule for value application, \Tvalapp, checks that the argument
$\lambda'$ is within the interval $(\Lambda_1,\Lambda_2)$, as required
by the parameter $\lambda$.


The rules in Figure~\ref{fig:static-instructions-pool} capture the
policy for lock usage.
Rule $\Tdone$ requires the release of all locks before terminating the
thread.
Rule $\Tfork$ splits permissions into sets $\Lambda$ and $\Lambda'$:
the former is transferred to the forked thread according to the
permissions required by the target code block, the latter remains with
the processor.
%
%
Rule $\Tnewlock$ assigns a lock type $\tupletype{\locktype}\lambda$ to
the register.
The new singleton lock type~$\lambda$ is recorded in
$\Psi$, so that it may be used in the rest of the instructions~$I$.
Rule $\Ttsl$ requires that the value under test is a lock in the heap
(of type~$\tupletype{\locktype}\lambda$) and records the type of the
lock value~$\lambda$ in register~$r$.
This rule also disallows testing a lock already held by the processor.
Rule $\Tunlock$ makes sure that only held locks are unlocked.
Rule $\TbranchC$ ensures that the processor holds the permission
required by the target code block, including the lock under test.
A processor is guaranteed to hold the tested lock only after
(conditionally) jumping to the critical region.
A previous test-and-set-lock instructions may have obtained the lock,
but the type system records that the processor holds the lock only
after the conditional jump.
The rule checks that the newly acquired lock is larger than all locks
in the possession of the thread.


The typing rules for memory and control flow are depicted in
Figure~\ref{fig:static-instructions-memory}.
Operations for loading from~($\Tload$), and for storing
into~($\Tstore$), tuples require that the processor holds the right
permissions (the locks for the tuples it reads from, or writes to).
Both rules preclude the direct manipulation of lock values by
programs, via the $\tau_n\neq\lambda'$ assumptions.


The rules for typing machine states are illustrated in
Figure~\ref{fig:static-machine}.
The rule for a thread item in the thread pool checks that the type and
required registers~$R$ are as expected in the type of the code block
pointed by~$v$.
Similarly, the rule for type checking a processor also permits that
type~$\Gamma$ of the registers~$R$ be more specific than the register
file type~$\Gamma'$ required to type check the remaining
instructions~$I$.
The heap value rule for code blocks adds to $\Psi$ each singleton lock
type (together with its bounds), so that they may be used in the rest
of the instructions~$I$.

\myparagraph{Example}
As expected, the example is not typable with the annotations
introduced previously. The three \lstinline|newLock| instructions
place in $\Psi$ three entries $f_1\colon(\emptyset, \emptyset)$,
$f_2\colon(\{f_1\},\{f_3\})$,
$f_3\colon(\{f_1\},\emptyset)$. Then the value
\lstinline|(liftLeftFork[f2])[f1]|
(in the example: \lstinline|liftLeftFork[f1,f2]|)
in the first \lstinline|fork| instruction issues goals $\Psi\vdash
\emptyset \prec f_1 \prec \emptyset$ and $\Psi\vdash \{f_1\} \prec f_2
\prec \emptyset$, which are easy to guarantee given that $\Psi$
contains an entry $f_2\colon(\{f_1\},\{f_3\})$.
Likewise, the second \lstinline|fork| instruction, generates goals
$\Psi\vdash \emptyset \prec f_2 \prec \emptyset$ and $\Psi\vdash \{f_2\}
\prec f_3 \prec \emptyset$, which are again hold because of same entry.
However, the last \lstinline|fork| instruction requires $\Psi\vdash
\emptyset \prec f_3 \prec \emptyset$ and $\Psi\vdash \{f_3\} \prec f_2
\prec \emptyset$, the second of which does not hold.

Notice however that each of the three \lstinline|jump| instructions
are typable per se.  For example, in code block
\lstinline|liftRightFork|, instruction
\lstinline|if r3 = 0b jump eat[l,m]| requires $\Psi \vdash \{l\}
\prec m$, which holds because  the signature for the code block
includes the annotation $m\colon(\{l\},\emptyset)$.

\myparagraph{Typable States Do Not Deadlock}
The main result of the type system, namely that $\Psi \vdash S$ and
$S\rightarrow^* S'$ implies $S'$ not deadlocked, follows from Subject
Reduction and from Typable States Are Not Deadlocked, in a
conventional manner.


\begin{lem}[Substitution Lemma]
  If $\Psi,\lambda\colon(\Lambda_1,\Lambda_2); \Gamma,
  \Lambda \vdash I$ and $\Psi(\lambda)'= (\Lambda_1,\Lambda_2)$,
  then $\Psi; \Gamma\sigma, \Lambda\sigma \vdash I\sigma$, where
  $\sigma= \subs{\lambda'}{\lambda}$.
\end{lem}


\begin{thm}[Subject Reduction]
  \label{thm:sr}
  If $\Psi \vdash S$ and $S\rightarrow S'$, then $\Psi' \vdash S'$,
  where $\Psi' = \Psi$ or $\Psi' =
  \Psi,l\colon\heaptuple{\pol\tau}{\lambda}$ (with $l$ fresh) or
  $\Psi' = \Psi,l\colon \heaptuple{\lambda}{\lambda},
  \lambda\colon(\pol\Lambda_1,\pol\Lambda_2)$ (with $l,\lambda$
  fresh).
\end{thm}
\begin{proof}
  (Outline) By induction on the derivation of $S \rightarrow S'$ proceeding by case analysis 
  on the last rule of the derivation, using the substitution lemma for rules \Rschedule, 
  \Rfork, \Rjump, and \RbranchT, as well as weakening in several rules.
\end{proof}

\begin{thm}[Typable States Are Not Deadlocked]
  \label{thm:ts}
  If $\Psi \vdash S$, then $S$ is not deadlocked.
\end{thm}
\begin{proof}
  (Sketch) Consider the contra-positive and show that deadlocked states are not
  typable. Without loss of generality suppose that $S$ is of the form
  $\program{H}{\langle t_{d_0},\dots,t_{d_m}\rangle}{\{d_{m+1}\colon
    p_{d_{m+1}}, \dots, d_n\colon p_{d_n}\}}$ with suspended threads
  $t_{d_i}$ and processors $p_{d_j}$ not necessarily distinct.

  Each of these threads and processors are trying to enter a critical
  region.
  For a processor $p_{d_i}$ we have that $S \rightarrow^*_{d_i} S'$
  where $d_i$-th processor in $S'$ is of the form
  $\processor{R}{\Lambda}{(\cjumpli;\_)}$ and $R(r) =
  \heaptuple{\_}{\lambda^{d_{i+1}}}$. By Subject Reduction $\Psi'
  \vdash S'$ where $\Psi'$ extends $\Psi$ as stated in
  Theorem~\ref{thm:sr}. A simple derivation starting from rule
  \TbranchC allows to conclude that $\Psi \vdash \Lambda_{d_i} \prec
  \lambda_{d_{i+1}}$.
  For a thread $t_{d_i} = \poolitem{l[\pol\lambda']}{R}$ in thread pool
  of~$S$ we run the machine $S''$ obtained from $S$ by replacing
  processor 1 with $\processor{R}{\Lambda}{I}
  \subs{\pol\lambda'}{\pol\lambda}$, where $H(l) = \heapcode
  {\codetypeforall {\pol\lambda\colon\_} {\_} {\Lambda}}
  {I}$. Proceeding as for processes above, we conclude again that
  $\Psi \vdash \Lambda_{d_i} \prec \lambda_{d_{i+1}}$.

  We thus have $\Psi \vdash \Lambda_{d_0} \prec \lambda_{d_1}$, \dots
  $\Psi \vdash \Lambda_{d_{n-1}} \prec \lambda_{d_n}$, $\Psi \vdash
  \Lambda_{d_n} \prec \lambda_{d_0}$ which is not satisfiable.
\end{proof}


\section{Type Inference}
\label{sec:type-inference}

Annotating lock ordering on large assembly programs may not be an easy
task.  In our setting, programmers (compilers, more often) produce
annotation free programs such as the one in
Figure~\ref{fig:philosophers}, and use an inference algorithm to
provide for the missing annotations.

\myparagraph{The Algorithm}
The \emph{annotation-free syntax} is obtained from that in
Figure~\ref{fig:syntax-instructions}, by removing the
$\colon(\Lambda,\Lambda)$ part both in the \newlock{} instruction and
in the universal type.
Given an annotation-free program~$H$, algorithm $\mathcal W$ produces
a pair, comprising a typing environment $\Psi$ and an annotated
program~$H^\star$, such that $\Psi \vdash H^\star$, or else fails.
In the former case $H^\star$ is typable, hence does not deadlock
(Theorem~\ref{thm:sr}); in the latter case, there is no possible
labeling for~$H$.

%
We depend on a set of \emph{variables over permissions} (sets of
locks), ranged over by~$\nu$, disjoint from the set of heap labels and
from the set singleton lock types introduced in
Section~\ref{sec:syntax}.
Constraints are computed by an intermediate step in our algorithm.


\begin{defi}[Constraints and solutions]~
  \begin{itemize}
  \item We consider \emph{constraints} of three distinct forms:
    $\Lambda \prec \lambda$, $\nu \prec \lambda$, and $\lambda \prec
    \nu$, and denote by $C$ a set of constraints;
  \item A \emph{substitution $\theta$} is a map from permission
    variables $\nu$ to permissions $\Lambda$;
  \item A substitution $\theta$ \emph{solves} $(\Psi, C)$ if
    $\Psi\theta \vdash x\theta \prec y\theta$ for all $x\prec y \in
    C$.
  \end{itemize}
\end{defi}

Algorithm $\mathcal W$ runs in two phases: the first, $\mathcal A$,
produces a triple comprising a typing environment~$\Psi$, an annotated
program~$H^\star$, and a collection of constraints $C$, all containing
variables over permissions $\Lambda$. The set of constraints is then
passed to a constraint solver, that either produces a
substitution~$\theta$ or fails.
In the former case, the output of $\mathcal W$ is the pair
$(\Psi\theta, H^\star\theta)$; in the latter $\mathcal W$ fails.
In practice, we do not need to generate $H^\star$ or to perform the
substitutions; our compiler accepts $H$ if the produced collection of
constraints is solvable, and rejects it otherwise.

\begin{myfigure}
  \begin{align*}
    &\tagA[ \{l_i \colon \heapcode {\tau_i} {I_i}\}_{i\in I}] =
    (\cup_{i\in I}\Psi_i, \{l_i \colon
    \heapcode {{\tau_i^\star}} {I_i^\star}
    \}_{i\in I}, \cup_{i\in I} C_i)
    \\
    &\quad\text{where }
    \tau_i^\star = \codetypeforall{\pol\lambda_i {\colon
        (\pol{\nu_i},\pol{\rho_i})}}{\Gamma_i}{\Lambda_i} = \tagT {\tau_i}
   \\
    &\quad\text{and }
    (I_i^\star, \Psi_i , C_i) = \mathcal I(I_i ,
    \{l_i\colon  {\tau_i^\star}\}_{i\in I} \cup\{\pol\lambda_i\colon
      (\pol{\nu_i},\pol{\rho_i})\},\Gamma_i,\Lambda_i)
   \\
    &\tagId {(\lambda,r \assign \newlock ;I)}  =
    ((\lambda{\colon(\nu,\rho)},r 
    \assign \newlock; I^\star), \Psi', C)
    \\
    &\quad \text{where}\;\;
    (I^\star, \Psi', C)  = \tagI I {\Psi\uplus\{\lambda{\colon(\nu,\rho)}\}} 
    {\Gamma\update{r}{\tupletype{\lambda}{\lambda}}} \Lambda
    \\
    &\tagId {(\cjumpli; I)}   =
    ((\cjumpli; I^\star), \Psi', C_1 \cup C_2 \cup \{{\Lambda \prec \lambda}\})
    \\
    &\quad\text{where }
    (\codetype{\Gamma'}{(\Lambda\uplus\{\lambda\})}, C_1) = \tagV{v}{\Psi}{\Gamma}
    \\
    &\quad\text{and } (I^\star, \Psi', C_2)  = \tagId I
    \\
    &\quad\text{and }
    \Psi \vdash \Gamma \subtype \Gamma'
    \\
    &\quad\text{and } \lambda = \Gamma(r)
    \\
   &\tagI {(\forki; I)} \Psi \Gamma {\Lambda} =
      ((\forki;I^\star), \Psi',C_1 \cup C_2)\\
    &\quad\text{where }
    (\codetype{\Gamma'}{\Lambda'}, C_1) = \tagV{v}{\Psi'}{\Gamma}
    \\
    &\quad\text{and }
    (I^\star, \Psi', C_2) = \tagI I {\Psi} {\Gamma} {\Lambda\setminus\Lambda'}
    \\
    &\quad\text{and }
    \Psi \vdash \Gamma \subtype \Gamma'
    \\
    &\tagV {\instantiation {\lambda} } \Psi \Gamma =
    (
    \tau
    \subs{\lambda}{\lambda'}, C \cup \{{\nu
      \prec \lambda \prec \rho}\})
    \\
    &\quad\text{where } (
    \forall[\lambda'\colon(\nu,\rho)]\tau,
    C) = \tagVd
    \\
    & \tagT { \codetypeforall {\pol\lambda} {(\register
        1\colon\tau_1,...,\register n\colon\tau_n)} {\Lambda} } =
    \codetypeforall { \tagLock{\pol\lambda}{\pol\nu}{\pol\rho} }
    {(\register 1\colon\tagT{\tau_1},...,\register
      n\colon\tagT{\tau_n})} {\Lambda}
  \end{align*}
  \caption{The tagging algorithm (selected rules).}
  \label{fig:algorithm}
\end{myfigure}


\myparagraph{Generating constraints}
Algorithm $\mathcal A$, described in Figure~\ref{fig:algorithm},
visits the program twice.
On a first step it builds an initial type environment $\Psi_0 =
\{l_i\colon {\tau_i^\star}\}_{i\in I}$ collecting the types for all
code blocks in the given program $\{l_i \colon \heapcode {\tau_i}
{I_i}\}_{i\in I}$, annotating with permission variables (denoted by
$\nu$ and $\rho$) the intervals for the locks bound in forall types;
on a second visit it generates the constraints and the annotated
syntax for the instructions in each code block.

The algorithm for instructions, $\mathcal I$, also shown in
Figure~\ref{fig:algorithm}, generates annotations for the singleton
lock type introduced in \lstinline|newLock| instructions, or further
constraints in the case of the jump-to-critical instruction.
In the case of a fork instruction, the algorithm calls function
$\mathcal V$ to obtain the required permission $\Lambda'$ and passes
the difference $\Lambda\setminus\Lambda'$ to the function that
annotates the continuation $I$.

The algorithm for values, $\mathcal V$, generates constraints in the
case of type application.
Finally, the algorithm for types annotates the singleton lock types in
forall types.
In the definition of all algorithms, permission-variables
$\nu,\rho,\pol\nu_i,\pol\rho_i$ are freshly introduced.

\myparagraph{Example}
For the running example, we first rename all bound variables so that
the type of code block \lstinline|liftLeftFork| mentions
\lstinline|l1| and \lstinline|m1|, that of \lstinline|liftLeftFork|
mentions \lstinline|l2| and \lstinline|m2|, and that of
\lstinline|eat| uses \lstinline|l3| and \lstinline|m3|.
For example:
\begin{lstlisting}
liftRightFork forall[l2].forall[m2].(r1:<l2>^L2, r2:<m2>^M2) requires {l2}
\end{lstlisting}

Then, algorithm $\mathcal A$ creates an initial environment $\Psi_0$
by generating twelve variables ($\rho_1$ to $\rho_{12}$) to annotate
the six locks ($\mathsf{l_i}$ and $\mathsf{m_i}$) in the three code
blocks that mention locks (\lstinline|liftLeftFork|,
\lstinline|liftRightFork|, and \lstinline|eat|). They are
$\mathsf{l_1}\colon(\rho_1,\rho_2)$, \dots,
$\mathsf{m_3}\colon(\rho_{11},\rho_{12})$.
Revisiting the signature of code block~\lstinline|liftRightFork|, we get:
\begin{lstlisting}
liftRightFork forall[m2::(rho7,rho8)].forall[l2::(rho5,rho6)].(r1:<l2>^L2, r2:<m2>^M2) requires {l2}
\end{lstlisting}

In the second pass, while in code block \lstinline|main|, algorithm
$\mathcal I$ generates six more permission variables ($\rho_{13}$ to
$\rho_{18}$) to annotate the new lock variables $\mathsf{f_1}$ to
$\mathsf{f_3}$ introduced with the \lstinline|newLock|
instructions. They are: $\mathsf{f_1}\colon(\rho_{13},\rho_{14})$
\dots $\mathsf{f_3}\colon(\rho_{17},\rho_{18})$.
The rest of the second pass generates new constraints in type
application and in jump-to-critical instructions. For example, in code
block \lstinline|liftRightFork|, and for value
\lstinline|eat[l2,m2]|, four constraints are generated: $\rho_9\prec
\mathsf{l_2}\prec \rho_{10}, \rho_{11} \prec \mathsf{m_2} \prec
\rho_{12}$. Then, in the jump-to-critical instruction,
\lstinline|if r3 = 0 jump eat[l2,m2]|, and since the thread holds lock
\lstinline|l2| (as witnessed by its signature
\lstinline|requires (l2)|), a new constraint $\{\mathsf{l_2}\} \prec
\mathsf{m_2}$ is generated.
The thus created set of constraints is then passed to a constraint
solver, which is bound to fail.

\myparagraph{Main result}
For soundness we start with a few lemmas.

\begin{lem}[Value soundness]
  \label{lem:value-soundness}
  If $\tagVd = (\tau, C)$ and $\theta$ solves $(\Psi,C)$ then
  $\Psi\theta; \Gamma\theta \vdash v\colon \tau\theta$.
\end{lem}

\begin{proof}
  (Outline) The proof proceeds by induction on the inference tree for
  $\Psi\theta; \Gamma\theta \vdash v\colon \tau\theta$ performing case
  analysis on the last typing rule applied.
\end{proof}

\begin{lem}[Instruction soundness] 
  \label{lem:instr-soundness}
  If $\tagId I = (I^\star, \Psi', C)$ and $\theta$ solves $(\Psi',C)$ then
  $\Psi'\theta; \Gamma\theta; \Lambda\theta \vdash I^\star\theta$ and 
  $\Psi \subseteq \Psi'$.
\end{lem}

\begin{proof}
  (Outline) The proof proceeds by induction on $I$. The cases for
  conditional jump and fork use Lemma~\ref{lem:value-soundness}.
\end{proof}

\begin{thm}[Soundness]
  \label{thm:soundness}
  If $\mathcal W(H) = (\Psi, H^\star)$ then $\Psi \vdash H^\star$. 
\end{thm}

\begin{proof}
  (Outline) Follows directly from Lemma~\ref{lem:instr-soundness}
  using typing rules for heap values and heaps. We use weakening on
  typing environments before applying the heap rule.
\end{proof}

Conversely, we believe that if $\Psi \vdash H^\star$,
then $\mathcal W(\erase(H^\star))$ does not fail, where $\erase$ is
the obvious lock-order annotation erasure function. A stronger result
would include a notion of principal solutions.




\section{Related Work and Conclusion}
\label{sec:conclusion}

\myparagraph{Related work}
The literature on type systems for deadlock freedom in lock-based
languages is vast; space restrictions prohibit a general survey.  We
however believe that the problem of type inference for deadlock
freedom in lock-based languages has been given not so much attention
in high-level languages, let alone low-level (assembly) languages.
Three characteristics separate our work from most proposals on the
topic: the non block structure of the locking primitives, the facts
that threads never block and that they may be suspended while holding
locks.


Following Coffman \etal one can classify the problem of deadlock under
the categories of \emph{detection and recovery},
\emph{avoidance} and \emph{prevention }~\cite{DBLP:conf/ictac/Boudol09,coffman.etal:system-deadlocks}.
In the first category, detection and recovery, on finds for example
works that check deadlocks at runtime.
Cunningham \etal infer locks for atomicity in an object-oriented
language, but use a runtime mechanism to detect when a thread's lock
acquisition would cause a
deadlock~\cite{cunningham.etal:lock-inference-proven}.
Java PathFinder~\cite{brat.etal:javapathfinder} and Driver
Verifier~\cite{ball.elal:thorough-static-analysis} identify violations
of the lock discipline during runtime tests. Agarwal
\etal~\cite{agarwal.etal:run-time-detection-deadlocks,agarwal.etal:detecting-deadlocks}
present an algorithm that detects potential deadlocks involving any
number of threads.
%

Under the \emph{avoidance} category on finds, e.g., a recent work by
Boudol where a type and effect system allows for the design of an
operational semantics that refuses to lock a pointer whenever it
anticipates to take a pointer that is held by another
thread~\cite{DBLP:conf/ictac/Boudol09}.


Our work falls into the third category above, \emph{prevention}.
Flanagan and Abadi present a functional language with mutable
references where locking is block structured and threads physically
block~\cite{flanagan.abadi:types-safe-locking}. From this work we
borrowed the idea of singleton lock types to describe, at the type
level, a single lock.
Type based deadlock prevention has also been study in the realm of
object-oriented languages, where, e.g., Boyapati \etal use a variant
of ownership types for preventing deadlocks in Java, performing
partial inference of annotations, but not of those related to lock
order~\cite{boyapati.lee.rinard:preventing-data-races}.

Suenaga proposes a concurrent functional language similar to Flanagan
and Abadi's mentioned above, except that it features non block
structured locking~\cite{DBLP:conf/aplas/Suenaga08}; his language
includes separate primitives for locking/unlocking, as in our case.
Albeit targeting at different level of abstraction, the results
(deadlock prevention) and the techniques (type inference) in both
works are similar, in particular the usage of a constraint-based
algorithm to infer types.
Differently from our case, Suenaga uses ownership types rather
than singleton lock types.
%

\myparagraph{Concluding remarks}
We have presented a type system that enforces a strict partial order
on lock acquisition, guaranteeing that well typed programs do not
deadlock. Towards this end we extended the syntax of our language to
incorporate annotations on the locking order. Acknowledging that the
annotation of large assembly programs (either manually or as the
result of a compilation process) is not plausible, we have introduced
an algorithm that infers the required annotations. The algorithm is
proved to be correct, hence that programs that pass our compiler are
exempt from deadlocks.

The current implementation of the algorithm generates, from a
non-annotated program, a set of constraints in the form of a Prolog
goal. The goal is then checked against a Prolog program that
implements the $\prec$ relation in Figure~\ref{fig:less-than}. We
consider the program typable if the goal succeeds. There is no point
in building the annotated syntax or performing the substitution, as
explained in Section~\ref{sec:type-inference}. Future work in this
area includes the automation of the whole process either by calling
the Prolog interpreter from within the compiler, or by implementing
relation $\prec$ directly in Java, the language of our type
checker/interpreter.

Future work also includes trying to assess the usage of our type
checker on larger programs, generated for example from an imperative
high-level language, and to further compare the singleton lock types
and the ownership types approaches for the description of non block
structured locking.


\bibliography{mil}
\bibliographystyle{eptcs}
\end{document}